\documentclass[journal]{IEEEtran}

\ifCLASSINFOpdf
\else
   \usepackage[dvips]{graphicx}
\fi
\usepackage{url}

\usepackage{graphicx}

\usepackage{amsmath,amsthm,amssymb,amsfonts, fancyhdr, color, comment, graphicx, environ}
\usepackage[monochrome]{xcolor}

\usepackage{multicol}
\usepackage{mdframed}
\usepackage{hhline}
\usepackage[shortlabels]{enumitem}
\usepackage{indentfirst}
\usepackage{diagbox}
\usepackage{mathtools}
\usepackage[super]{nth}
\usepackage{hyperref}
\usepackage{arydshln}
\usepackage{mathabx}
\usepackage{multicol}
\usepackage{changepage}
\usepackage[nolist,nohyperlinks]{acronym}

\DeclareMathOperator*{\argmax}{arg\,max}
\DeclareMathOperator*{\argmin}{arg\,min}

\renewcommand{\vec}[1]{\text{vec}\left(#1\right)}

\acrodef{UE}{user equipment}
\acrodef{BS}{base station}
\acrodef{Tx}{transmitter}
\acrodef{Rx}{receiver}
\acrodef{CFO}{carrier frequency offset}
\acrodef{ULA}{uniform linear array}
\acrodef{OFDM}{orthogonal frequency division multiplexing}
\acrodef{AWGN}{additive white Gaussian noise}
\acrodef{LO}{local oscillator}
\acrodef{MAP}{maximum \textit{a posteriori}}
\acrodef{mmWave}{millimeter wave}
\acrodef{subTHz}{sub-terahertz}
\acrodef{LOS}{line-of-sight}
\acrodef{NLOS}{non-line-of-sight}
\acrodef{MCMC}{Markov chain Monte Carlo}
\acrodef{VB}{variational Bayes}
\acrodef{EP}{expectation propagation}
\acrodef{KL}{Kullback-Leibler}
\acrodef{SBL}{sparse Bayesian learning}
\acrodef{RV}{random variable}
\acrodef{PDF}{probability density function}
\acrodef{EM}{expectation-maximization}
\acrodef{5G}{fifth generation}
\acrodef{iid}{independent identically distributed}
\acrodef{ML}{maximum likelihood}
\acrodef{SCA}{successive convex approximation}
\acrodef{MM}{majorization-minimization}
\acrodef{ADMM}{alternating direction method of multipliers}
\acrodef{SNR}{signal-to-noise ratio}
\acrodef{AoA}{angle of arrival}
\acrodefplural{AoA}[AoAs]{angles of arrival}
\acrodef{AoD}{angle of departure}
\acrodefplural{AoD}[AoDs]{angles of departure}
\acrodef{MSE}{mean squared error}
\acrodef{RMSE}{root mean squared error}
\acrodef{ReMSE}{relative mean squared error}
\acrodef{PSO}{particle swarm optimization}
\acrodef{NM}{Nelder-Mead}
\acrodef{B5G}{beyond 5G}
\acrodef{6G}{sixth generation standard}
\acrodef{MIMO}{multiple input multiple output}
\acrodef{LLN}{law of large numbers}
\acrodef{LGR}{large grid regime}
\acrodef{LPA}{long pilot approximation}
\acrodef{MC}{Monte Carlo}
\acrodef{ToF}{time of flight}
\acrodefplural{ToF}[ToFs]{times of flight}
\acrodefplural{TOF}[TOFs]{times of flight}
\acrodef{CACC}{cross-antenna cross-correlation}
\acrodef{MMSE}{minimum mean square error}
\acrodef{DTFT}{discrete-time Fourier transform}
\acrodef{DFT}{discrete Fourier transform}
\acrodef{FFT}{fast Fourier transform}
\acrodef{2P2T}{two precoders two transmissions}
\acrodef{TVP}{time-varying precoder}
\acrodef{RF}{radio frequency}
\acrodef{PSF}{point spread function}
\acrodef{DBSCAN}{density-based spatial clustering of applications with noise}
\acrodef{SSE}{Sum Square Error}
\acrodef{TDMA}{time division multiple access}
\acrodef{FDMA}{frequency division multiple access}
\acrodef{CDMA}{code division multiple access}
\acrodef{BER}{bit error rate}
\acrodef{MUSIC}{multiple signal classification}
\acrodef{URA}{uniform rectangular array}
\acrodef{JCS}{joint communication and sensing}
\acrodef{NPNT}{N-precoders N-transmissions}
\acrodef{SOR}{successive over-relaxation}
\acrodef{COMPAS}{concurrent mapping, positioning, and synchronization}
\acrodef{SLAM}{simultaneous localization and mapping}
\acrodef{AECD}{alternating exact coordinate descent}
\acrodef{ISAC}{integrated sensing and communications}
\acrodef{JSC}{joint sensing and communication}
\acrodef{FR2}{frequency range 2}
\acrodef{CSI}{channel state information}
\acrodef{SAGE}{space-alternating generalized expectation-maximization}
\acrodef{OR}{over-relaxation}
\acrodef{AIC}{Akaike information criterion}
\acrodef{SIMD}{single instruction multiple data}
\acrodef{ASIC}{application specific integrated circuit}
\acrodef{FPGA}{field-programmable gate array}
\acrodef{MAC}{multiply–accumulate}
\acrodef{CP}{canonical polyadic}
\acrodef{ESPRIT}{estimation of signal parameters via rotational invariance techniques}
\acrodef{MHR}{multidimensional harmonic retrieval}
\acrodef{i.i.d.}{independent identically distributed}
\acrodef{HST}{high-speed train}
\acrodef{ICI}{intercarrier interference}
\acrodef{DFRC}{dual-functional radar communications}
\acrodef{LS}{least squares}
\acrodef{MSVD}{multilinear singular value decomposition}
\acrodef{ALS}{alternating least squares}
\acrodef{dGN}{damped Gauss-Newton}
\acrodef{WSNSAP}{wideband spatial nonstationary wireless channels with antenna polarization}
\acrodef{GN}{Gauss-Newton}
\acrodef{RiMAX}{Richter's maximum likelihood estimation}
\acrodef{BIC}{Bayesian information criteria}
\acrodef{XL-MIMO}{extra-large MIMO}
\acrodef{FR2}{frequency range 2}
\acrodef{MDL}{minimum description length}
\acrodef{mMIMO}{massive MIMO}
\acrodef{NLS}{nonlinear least squares}
\acrodef{MAE}{mean absolute error}
\acrodef{MedAE}{median absolute error}
\acrodef{CDF}{cumulative distribution function}
\acrodef{SIC}{successive interference cancellation}
\acrodef{MC}{Monte Carlo}
\acrodef{QPSK}{quadrature phase-shift keying}
\acrodef{std-SAGE}{standard SAGE}
\acrodef{GLRT}{generalized likelihood ratio test}
\acrodef{DOF}{degree of freedom}
\acrodef{PCE}{parametric channel estimation}
\acrodef{ADC}{analog to digital converter}
\acrodef{LNA}{low noise amplifier}
\acrodef{LM}{Lloyd-Max}
\acrodef{DAC}{digital to analog converter}
\acrodef{PA}{power amplifier}
\acrodef{ACD}{alternating coordinate descent}
\acrodef{GEVD}{generalized eigenvalue decomposition}
\acrodef{BFGS}{Broyden–Fletcher–Goldfarb–Shanno}
\acrodef{GN-DL}{Gauss--Newton with dogleg trust region}
\acrodef{SVD}{singular value decomposition}
\acrodef{EVD}{eigenvalue decomposition}
\acrodef{JADE}{joint AoD and Doppler shift estimation}
\acrodef{CREL-CP}{component rescaling with exact line search CP}
\acrodef{CP-TVFS}{CP time-varying and frequency-selective}
\acrodef{MS-SAGE}{multiple-start SAGE}
\acrodef{SOTA}{state-of-the-art}

\usepackage{stmaryrd}
\SetSymbolFont{stmry}{bold}{U}{stmry}{m}{n}

\usepackage{algorithm}
\usepackage[noend]{algpseudocode}

\newcommand{\secgap}{\vspace{-0.2cm}}
\makeatletter
\def\section{\@startsection{section}{1}{\z@}{3.0ex plus 0.3ex minus 1.5ex}%
{0.7ex plus 0.2ex minus 0ex}{\normalfont\normalsize\centering\scshape}}%
\def\subsection{\@startsection{subsection}{2}{\z@}{3.5ex plus 0.3ex minus 1.5ex}%
{0.7ex plus 0.2ex minus 0ex}{\normalfont\normalsize\itshape}}%
\makeatother
\AtBeginDocument{%
  \setlength{\abovedisplayskip}{2pt plus 2pt minus 1pt}%
  \setlength{\belowdisplayskip}{2pt plus 2pt minus 1pt}%
  \setlength{\abovedisplayshortskip}{1pt plus 1pt}%
  \setlength{\belowdisplayshortskip}{2pt plus 1pt}%
  \setlength{\jot}{2pt}}                                 

\begin{document}

\title{Time-Varying Parametric Channel Estimation With CP Decomposition Tensor Processing}

\author{Enrique T. R. Pinto, \IEEEmembership{Graduate Student Member, IEEE}, André L. F. de Almeida, \IEEEmembership{Senior Member, IEEE}, \mbox{and Markku Juntti}, \IEEEmembership{Fellow, IEEE}

\thanks{The work was supported in part by the Research Council of Finland (former Academy of Finland) 6G Flagship Program (Grant Number: 369116) and S6GRAN project (370559).}
\thanks{Enrique T. R. Pinto and Markku Juntti are with the Centre for Wireless Communications, Faculty
of Information Technology and Electrical Engineering, University of Oulu, 90014 Oulu, Finland (e-mail: \{enrique.pinto, markku.juntti\}@oulu.fi).}
\thanks{André L. F. de Almeida is with the Department of Teleinformatics Engineering, Federal University of Ceará, 60455-970 Fortaleza, Brazil (e-mail: andre@gtel.ufc.br).}
}

\maketitle

\begin{abstract}
Integrated sensing and communications (ISAC) is a key use case for sixth-generation (6G) wireless systems, where parametric channel estimation (PCE) plays a central role in enabling sensing, localization, and channel equalization in high-mobility scenarios. However, conventional estimation is linear in the pilot
observations, while PCE requires solving a nonlinear, nonconvex
multidimensional harmonic retrieval problem. This motivates the development of lower-complexity solutions. In this letter, we propose a fast PCE algorithm for time-varying and frequency-selective (TVFS) channels based on canonical polyadic (CP) decomposition and tensor processing, combined with ESPRIT-based initialization, component rescaling, and exact line-search alternating coordinate descent. Single-stream and multi-stream versions of the estimator are described. Numerical results show that the proposed method clearly outperforms a related CP-based baseline while achieving estimation performance close to a multiple-start space-alternating generalized expectation-maximization (SAGE) benchmark at a substantially lower computational cost, with much lower execution time.
\end{abstract}

\vspace{-0.2cm}

\IEEEpeerreviewmaketitle

\renewcommand{\baselinestretch}{.9}

\vspace{-0.45cm}
\section{Introduction}

\IEEEPARstart{T}{ensor} processing methods have garnered increasing interest in wireless communications because they preserve the multilinear structure present in many signal models \cite{tensors_wcom,parafac_unified_wcom,tst_coding_mimo,semi_blind_nested_parafac_coop_mimo,irs_tensor_channel_estimation}. In parallel, the emergence of 
\ac{ISAC} for the \ac{6G} has intensified the demand for accurate sensing, localization, and channel inference techniques \cite{isac1,isac2}. Among these, \ac{PCE} plays a central role in providing geometric channel information, including the number of paths and their \acp{AoA}, \acp{AoD}, \acp{ToF}, and Doppler frequencies \cite{zhang_enabling_jsc,henk_cp}. Despite this potential, most tensor-based approaches are not well suited to time-varying scenarios: they either neglect Doppler, assume constant channels over subframes, or impose restrictive pilot structures.
\par

Tensor-based methods have been successfully applied to blind multiuser equalization \cite{parafac_unified_wcom}, space-time coding \cite{tst_coding_mimo}, cooperative and relay-assisted \ac{MIMO} systems \cite{semi_blind_nested_parafac_coop_mimo}, and intelligent reflecting surface aided channels \cite{irs_tensor_channel_estimation}; see also \cite{tensors_wcom,Favier_EUSIPICO,Sidiropoulos_2017} for a broad overview. More specifically, \cite{tensor_ch_est_massive_mimo_ofdm} proposed a tensor-based compressive sensing framework for massive \ac{MIMO} \ac{OFDM} channel estimation by exploiting the multilinear sparse structure of time- and frequency-selective channels. Cao \textit{et al.} \cite{hyb_field_tensor} proposed a hybrid-field channel estimation method based on the \ac{CP} decomposition, while Wen \textit{et al.}  \cite{henk_cp} developed a \ac{CP}- and tensor-\ac{ESPRIT}-based method for \ac{PCE} and mapping with diffuse multipath. In \cite{rankaware_completion}, rank-aware matrix completion targets time-varying channels using on-grid estimation. However, these methods do not explicitly address Doppler shift estimation. Moreover, methods such as \cite{doppler_two_stage1, doppler_two_stage2} require a two-stage pilot transmission, thereby increasing estimation overhead, and assume that the number of paths is known \textit{a priori}, which is unrealistic in practice. The method in \cite{cp-tvfs} estimates most channel parameters, but its Doppler estimation uses the current path magnitude estimate, and the phase error propagates into the Doppler estimate.

The main appeal of tensor-based \ac{PCE} methods lies in their computational efficiency. Once the \ac{CP} components have been estimated, several subsequent computations can be carried out in parallel. This property is particularly attractive in large-scale antenna arrays and high-mobility scenarios, where the channel model dimension increases significantly and low-latency processing becomes essential. In addition, multilinear factorization offers a compact representation of the received data, which helps separate the contributions of different channel parameters before the final refinement stage. This contrasts with \ac{ML} methods based on \ac{EM} \cite{EM} such as \ac{SAGE} \cite{sage,vb_sage,ch_est_pinto} and RiMAX \cite{rimax}, which are more sequential and have more computationally demanding parameter updates. Although tensor-\ac{ESPRIT} methods can jointly recover a \ac{CP} decomposition and the generators of the Vandermonde factors \cite{sorensen_tensor_esprit}, they are not directly applicable here because the time-varying precoder-weighted factor in the single-stream model is generally not Vandermonde, while hybrid precoding and combining further mix the transmit and receive steering vectors. 

In this letter, we address this gap by proposing a single-stage, tensor-based \ac{PCE} procedure that explicitly estimates Doppler with low computational complexity. The proposed method, named \ac{CREL-CP}, combines \ac{CP} decomposition, \ac{ESPRIT}, rescaling, and exact line-search \ac{ACD} in a unified estimation pipeline. Its originality lies in coupling a low-complexity multilinear factorization stage with subsequent grid-free estimation and component rescaling to resolve the CP scaling ambiguity, enabling accurate path-gain and channel reconstruction. We further present two tailored versions of the method: one for single-stream pilots with a fully digital receiver and another for multi-stream pilots with a hybrid combiner. Overall, the proposed framework aims to provide a fast \ac{PCE} solution with competitive estimation accuracy and model-order recovery.

\secgap
\section{System Model} \label{sec:model}
Consider a \ac{Tx}-\ac{Rx} pair equipped with $\lambda/2$-spaced \acp{ULA}. An \ac{OFDM} pilot signal \mbox{$\mathcal{X}\in\mathbb{C}^{N_\text{C} \times N_\text{S} \times N_\text{T}}$} is transmitted over a fast time-varying channel $\mathcal{H}\in\mathbb{C}^{N_\text{C} \times N_\text{S} \times N_\text{R} \times N_\text{T}}$, where $N_\text{C}$, $N_\text{S}$, $N_\text{R}$, and $N_\text{T}$ denote the numbers of \ac{OFDM} subcarriers, \ac{OFDM} symbols, receive antennas, and transmit antennas, respectively. We assume a far-field narrowband (with respect to the array response, no beam squint) channel composed of the superposition of $L$ paths, each modeled as a rank-1 contribution, i.e.,
\begin{equation} \label{eq:channel_model}
    [\mathcal{H}]_{ntuv} = h_{ntuv} = \sum^L_{\ell=1} b_\ell e^{j(n\omega_{1\ell} + t\omega_{2\ell} + u\psi_\ell + v\varsigma_\ell)},
\end{equation}
where $b_\ell\in\mathbb{C}$ is the path coefficient, while \mbox{$\omega_{1\ell}=-2\pi f_{\mathrm{scs}}\tau_\ell$}, $\omega_{2\ell}=2\pi f_{D\ell}T_{\mathrm{sym}}$, $\psi_\ell=\pi\sin\theta_\ell^{\mathrm{AoA}}$, and $\varsigma_\ell=\pi\sin\theta_\ell^{\mathrm{AoD}}$ are the angular frequencies associated with the \ac{ToF}, Doppler, \ac{AoA}, and \ac{AoD}, respectively. Also, $f_{\mathrm{scs}}$, $\tau_\ell$, $f_{D\ell}$, $T_{\mathrm{sym}}$, $\theta_\ell^{\mathrm{AoA}}$, and $\theta_\ell^{\mathrm{AoD}}$ respectively stand for the subcarrier spacing, \ac{ToF}, Doppler shift, symbol time, \ac{AoA}, and \ac{AoD}.
Moreover, $[\mathcal{H}]_{ntuv}$ denotes the entry of tensor $\mathcal{H}$ indexed by $(n,t,u,v)$, where the indices correspond to subcarrier, \ac{OFDM} symbol, receive antenna, and transmit antenna, respectively. The model assumes $|\omega_{2\ell}| \ll \pi$, i.e., \ac{ICI} is negligible. This is approximately satisfied in many use cases even in \ac{mmWave}.

\secgap
\subsection{Single-Stream Pilot With Fully Digital Receiver} \label{subsec:model_ss}
For a fully digital receiver, the received signal is represented by the tensor $\mathcal{Y}$, with entries given by
\begin{equation}
    [\mathcal{Y}]_{ntu} = y_{ntu} = \sum^{N_\text{T}-1}_{v=0} h_{ntuv} x_{ntv} + w_{ntu},
\end{equation}
where $w_{ntu}\sim\mathcal{CN}(0,N_0)$ are the entries of the circularly symmetric \ac{AWGN} tensor $\mathcal{W}$. Assume that the pilot signal admits the separable structure $x_{ntv} = p_{tv}s_{nt}$, i.e., the product of a time-varying precoder and an \ac{OFDM} pilot resource grid. \textcolor{blue}{Assume further that $s_{nt} = s_t$, i.e., it is a rank-1 matrix constant over $n$}\footnote{\textcolor{blue}{This preserves the \ac{CP} structure of $\mathcal{A}$ under \ac{ICI}, i.e., $\mathcal{A}$ can still be written as the sum of rank-1 tensors, and keeps the subcarrier factor of each path nearly exponential, limiting the ICI-induced error in the delay estimate $\hat{\omega}_{1\ell}$.}}. Then, assuming constant-magnitude pilot symbols, we define the tensor $\mathcal{A}$ as
\begin{multline}\label{eq:tensor_A}
    [\mathcal{A}]_{ntu} = a_{ntu} = \frac{y_{ntu}}{s_{nt}}\\
    =  \sum^L_{\ell=1} e^{jn\omega_{1\ell}} \left(\sum_v b_\ell e^{j(t\omega_{2\ell} + v\varsigma_\ell)} p_{tv}\right) e^{ju\psi_\ell} + \frac{w_{ntu}}{s_{nt}}.
\end{multline}
The tensor $\mathcal{A}$ is the main observation used for parameter estimation in Section \ref{sec:digital_estimation}. 

\secgap
\subsection{Multi-Stream Pilot With Hybrid Combiner}
We now extend the model to a hybrid beamforming architecture. For a \ac{Tx}-\ac{Rx} pair with hybrid beamformers, the received contribution associated with path $\ell$ is written as
\begin{equation} \label{eq:rx_sig}
    y^\ell_{ntm} = b_\ell r_{mt}(\psi_\ell) x_{nt}(\varsigma_\ell) e^{j(n \omega_{1\ell} + t\omega_{2\ell})},
\end{equation}
where
\begin{multicols}{2}\noindent
    \begin{equation} \label{eq:r_psi}
        r_{mt}(\psi_\ell) = \sum^{N_\text{R}-1}_{u=0} r_{mtu} e^{ju\psi_\ell},
    \end{equation}
    \begin{equation} \label{eq:x_sigma}
        x_{nt}(\varsigma_\ell) = \sum^{N_\text{T}-1}_{v=0} x_{ntv} e^{jv\varsigma_\ell},
    \end{equation}
\end{multicols}\noindent
with $r_{mtu}$ denoting the entries of the combiner tensor $\mathcal{R}$ at \ac{OFDM} symbol $t$, radio frequency chain \textcolor{blue}{\mbox{$m\in\{0,\dots,D_\text{R}-1\}$}}, and receive antenna $u$, and $x_{ntv}$ denoting the transmitted pilot entry at subcarrier~$n$, symbol $t$, and transmit antenna $v$. The complete received signal is then given by
\begin{equation} \label{eq:rx_sig_hybrid}
    y_{ntm} = \sum^{L}_{\ell=1} y^\ell_{ntm} + w_{ntm} = \sum^{N_\text{R}-1}_{u=0} \sum^{N_\text{T}-1}_{v=0} r_{mtu} x_{ntv} h_{ntuv} + w_{ntm},
\end{equation}
where the transmitted signal is generally decomposed into $D_\text{T}$ transmit streams as $[\mathcal{X}]_{ntv} = x_{ntv} = \sum^{D_\text{T}-1}_{d=0} p_{vdt} s_{ntd}$,
with $p_{vdt}$ and $s_{ntd}$ denoting the entries of the precoding tensor $\mathcal{P}$ and pilot-symbol tensor $\mathcal{S}$, respectively.

\secgap
\section{Mathematical Preliminaries}
In this section, we review the mathematical tools underlying the proposed estimation procedures: the \ac{CP} decomposition \cite{cp_original} and the \ac{MDL}
\cite{mdl_tensor}.

\secgap
\subsection{CP Decomposition} \label{sec:cp}
The proposed estimators make a recurrent use of the \ac{CP} decomposition. For a third-order tensor $\mathcal{Z} \in \mathbb{C}^{N_1\times N_2 \times N_3}$, its rank-$K$ \ac{CP} decomposition is written as
\begin{equation} \label{eq:cp_example}
    \mathcal{Z} = \llbracket \mathbf{Z}_1, \mathbf{Z}_2, \mathbf{Z}_3 \rrbracket = \sum^K_{k=1} \mathbf{z}^1_{k} \circ \mathbf{z}^2_{k} \circ \mathbf{z}^3_{k},
\end{equation}
where ``$\circ$'' denotes the tensor outer product. The factor matrices are defined as $\mathbf{Z}_m = \begin{bmatrix} \mathbf{z}^m_{1} & \cdots & \mathbf{z}^m_{K} \end{bmatrix}$, with $\mathbf{z}^m_{k} \in \mathbb{C}^{N_m}$. In our context, this decomposition is useful because the observed data tensors approximately follow the multilinear structure in \eqref{eq:cp_example}. A well-known property of the \ac{CP} decomposition is the scaling ambiguity among the factors, since only the product of the component scalings is identifiable. Therefore, different sets of factors are equivalent as long as they yield the same rank-1 tensor terms \textcolor{blue}{$\mathbf{z}^1_k\circ\mathbf{z}^2_k\circ\mathbf{z}^3_k$} for each $k$.

\secgap
\subsection{Model-Order Estimation} \label{sec:model_order_est}
To estimate the tensor rank $L$, which serves as an input to the \ac{CP} decomposition, we apply the \ac{MDL} criterion to all $k$-mode unfoldings of the input tensor $\mathcal{Z}$. Let $\mathcal{Z}_{(k)}$ denote the unfolding of $\mathcal{Z}$ along the $k$th mode \cite{tensorreview}. The resulting estimate of $L$ is defined as
\begin{equation} \label{eq:mdl_est}
    \hat{L} = \max_k \mathrm{MDL}\left(\mathcal{Z}_{(k)}\right),
\end{equation}
where $\mathrm{MDL}(\cdot)$ is evaluated as in \cite[Eq. (7)]{mdl_tensor}.

\secgap
\section{Estimation Procedure}
We now describe the proposed estimation procedures for the two modeling cases introduced in Section \ref{sec:model}: (a) a single-stream pilot with a fully digital receiver, and (b) a multi-stream pilot with a hybrid combiner. In both cases, the estimation strategy follows the same general sequence: (1) model-order estimation, (2) tensor decomposition, (3) estimation of directly identifiable parameters, (4) \textcolor{blue}{rescaling} of the tensor components, and (5) estimation of the remaining parameters.

\secgap
\subsection{Digital Array Channel Estimation} \label{sec:digital_estimation}
For the digital case, we use the single-stream pilot and process the tensor $\mathcal{A}$ defined in \eqref{eq:tensor_A}. We begin by estimating $\hat{L}$ as described in Section \ref{sec:model_order_est}, and then compute the rank-$\hat{L}$ \ac{CP} decomposition of $\mathcal{A}$. This yields components of the form
\begin{align}
    &[\mathbf{a}^1_\ell]_n \: \propto \: e^{jn\omega_{1\ell}} \\
    &[\mathbf{a}^2_\ell]_t \: \propto \: \sum_v b_\ell e^{ j ( t\omega_{2\ell} + v\varsigma_\ell )} p_{tv}, \label{eq:a2}\\
    &[\mathbf{a}^3_\ell]_u \: \propto \: e^{ju\psi_\ell},
\end{align}
where the noise contribution is omitted for simplicity. \textcolor{blue}{The vectors $\mathbf{a}^m_\ell$, $m=1,2,3$, are the CP factors, known up to the scaling ambiguity of Section~\ref{sec:cp}.} The factors $[\mathbf{a}^1_\ell]_n$ and $[\mathbf{a}^3_\ell]_u$ allow us to estimate $\omega_{1\ell}$ and $\psi_\ell$, respectively, via \ac{ESPRIT} as in \cite{henk_cp}. Using the estimates $\hat{\omega}_{1\ell}$ and $\hat{\psi}_{\ell}$, reconstruct the corresponding structured factors, denoted by \textcolor{blue}{the respective ``hatted'' equivalents} $\hat{\mathbf{a}}^1_\ell$ and $\hat{\mathbf{a}}^3_\ell$, as
\begin{align}
    \hat{\mathbf{a}}^1_\ell &= \begin{bmatrix} 1 & e^{j\hat{\omega}_{1\ell}} & \cdots & e^{j(N_\text{C}-1)\hat{\omega}_{1\ell}}\end{bmatrix}^T, \\
    \hat{\mathbf{a}}^3_\ell &= \begin{bmatrix} 1 & e^{j\hat{\psi}_{\ell}} & \cdots & e^{j(N_\text{R}-1)\hat{\psi}_{\ell}}\end{bmatrix}^T.
\end{align}
Next, consider the $\ell$th rank-1 component tensor $\mathcal{A}_\ell$. For convenience, we define $\mathbf{a}_\ell = \vec{\mathcal{A}_\ell} = \vec{\mathbf{a}^1_\ell \otimes \mathbf{a}^3_\ell \otimes \mathbf{a}^2_\ell}$,
where vectorization is taken columnwise.
\textcolor{blue}{The least-squares estimate of $\mathbf{a}_\ell^{2}$ given the reconstructed factors $\hat{\mathbf{a}}_\ell^{1}$ and $\hat{\mathbf{a}}_\ell^{3}$, $\hat{\mathbf{a}}_\ell^{2}=\arg\min_{\mathbf{a}^{2}_\ell} \big\|\mathbf{a}_\ell-\hat{\mathbf{a}}_\ell^{1}\otimes\hat{\mathbf{a}}_\ell^{3} \otimes\mathbf{a}^{2}_\ell\big\|^{2}$, has the closed-form}
\begin{equation} \label{eq:a2_refining}
    \hat{\mathbf{a}}^2_\ell = \frac{(\hat{\mathbf{a}}_\ell^{3H}\mathbf{a}^3_\ell)(\hat{\mathbf{a}}_\ell^{1H}\mathbf{a}^1_\ell)}{\|\hat{\mathbf{a}}_\ell^3\|^2\|\hat{\mathbf{a}}_\ell^1\|^2}\,\mathbf{a}^2_\ell,
\end{equation}
which corresponds to a rescaling of \textcolor{blue}{$\mathbf{a}^2_\ell$}. This allows us to accurately compute both the magnitude and phase of $b_\ell$. Thus, we can accurately reconstruct the wireless channel. Having obtained $\hat{\mathbf{a}}^2_\ell$, we estimate the remaining parameters by solving
\begin{gather} \label{eq:min_prob_omega_sigma}
    (\hat{\omega}_{2\ell}, \hat{\varsigma}_\ell, \hat{b}_\ell ) = \argmin_{(\omega_2,\varsigma,b)} \left\| \hat{\mathbf{a}}^2_\ell - b\boldsymbol{\alpha}(\omega_2,\varsigma) \right\|^2, \\
    [\boldsymbol{\alpha}(\omega_2,\varsigma)]_t = \sum_v e^{j(t\omega_{2} + v\varsigma)} p_{tv}.
\end{gather}
For fixed $(\omega_{2}, \varsigma)$, the least-squares estimate of $b_\ell$ has the closed-form expression
\begin{equation} \label{eq:b_expression}
    b_\ell(\omega_{2}, \varsigma) = \hat{\mathbf{a}}^{2,T}_\ell \boldsymbol{\alpha}^*(\omega_2,\varsigma)/ \| \boldsymbol{\alpha}(\omega_2,\varsigma) \|^2.
\end{equation}
Substituting this expression into \eqref{eq:min_prob_omega_sigma} yields an equivalent optimization problem over $(\omega_{2\ell}, \varsigma_\ell)$ only,
\begin{equation} \label{eq:omega2_varsigma_est}
    (\hat{\omega}_{2\ell}, \hat{\varsigma}_\ell) = \argmax_{(\omega_{2}, \varsigma)} \frac{\left| \hat{\mathbf{a}}^{2,T}_\ell \boldsymbol{\alpha}^*(\omega_2,\varsigma) \right|^2}{\sum_t \left|\sum_v e^{jv\varsigma} p_{tv} \right|^2 }.
\end{equation}
This optimization is referred to in the literature as \ac{JADE} \cite{cp-tvfs}. Since an exhaustive grid search is computationally demanding, we instead adopt the \ac{ACD} strategy with exact line search from \cite{ch_est_pinto}; a multiple-start heuristic may also be used to mitigate convergence to poor local optima. Once $(\hat{\omega}_{2\ell}, \hat{\varsigma}_\ell)$ are obtained, we compute $\hat{b}_\ell = b_\ell(\hat{\omega}_{2\ell}, \hat{\varsigma}_\ell)$. Repeating this procedure for $\ell = 1,\dots,\hat{L}$ yields the full set of parameter estimates $\hat{\boldsymbol{\xi}}=( \hat{\boldsymbol{\xi}}_1, \dots, \hat{\boldsymbol{\xi}}_{\hat{L}} )$, where $\hat{\boldsymbol{\xi}}_\ell = (\hat{\psi}_\ell, \hat{\varsigma}_\ell, \hat{\omega}_{1\ell}, \hat{\omega}_{2\ell}, \hat{b}_\ell)$. Algorithm~\ref{alg:main_ss} summarizes the proposed estimation procedure.

\begin{algorithm}[!t]
\caption{Proposed single-stream \ac{CREL-CP} algorithm}
\label{alg:main_ss}
\begin{algorithmic}[1]
    \Procedure{Main}{$\mathcal{Y}$, $\mathcal{X}$}
    \State Compute $\mathcal{A}$ as in (\ref{eq:tensor_A}); 
    \State Compute $\hat{L}$ using (\ref{eq:mdl_est}) on $\mathcal{A}$;
    \State Compute the \ac{CP} decomposition of $\mathcal{A}$;
    \For{$\ell = 1,\,\dots,\,\hat{L}$}
        \State Estimate $\omega_{1\ell}$ using \ac{ESPRIT};
        \State Estimate $\psi_{\ell}$ using \ac{ESPRIT};
        \State Resolve the component scaling of $\mathbf{a}^2_\ell$ as in (\ref{eq:a2_refining}); 
        \State Solve (\ref{eq:omega2_varsigma_est}) using the \ac{ACD} procedure with exact line search from \cite{ch_est_pinto}; 
        \State Compute $b_\ell(\hat{\omega}_{2\ell},\hat{\varsigma}_\ell)$ using (\ref{eq:b_expression});
        \State Group the estimates as $\hat{\boldsymbol{\xi}}_\ell = (\hat{\psi}_\ell, \hat{\varsigma}_\ell, \hat{\omega}_{1\ell}, \hat{\omega}_{2\ell}, \hat{b}_\ell)$;
    \EndFor
    \Return $\hat{\boldsymbol{\xi}} = (\hat{\boldsymbol{\xi}}_1,\dots, \hat{\boldsymbol{\xi}}_{\hat{L}})$, $\hat{L}$
\EndProcedure
\end{algorithmic}
\end{algorithm}

\secgap
\subsection{Hybrid Array Channel Estimation}
We next extend the procedure to the hybrid beamforming case. Consider a \ac{Tx}-\ac{Rx} pair equipped with hybrid beamformers, so that the received signal tensor $\mathcal{Y}$ has entries $y_{ntm}$ given by \eqref{eq:rx_sig_hybrid}. We assume partially connected \cite[Fig.~1]{hw_impaired_ch_est} hybrid beamformers with \ac{Tx} and \ac{Rx} subarray sizes given by $N^\text{T}_a$ and $N^\text{R}_a$, respectively. Because multiple pilot streams can be transmitted simultaneously, we can no longer divide by $s_{nt}$ to decouple the $n$ and $t$ dimensions as in the construction of $\mathcal{A}$. To recover a tractable multilinear structure, we therefore assume a pilot and combiner that are constant over $t$, so that the entries of $\mathcal{X}$ and $\mathcal{R}$ can be written as $x_{nv} = \sum^{D_\text{T}-1}_{d=0} p_{vd} s_{nd}$ and $r_{mu}$, respectively. Under this assumption, we have
\begin{equation}
     y_{ntm} = \sum^L_{\ell=1} b_\ell x_{n}(\varsigma_\ell) e^{j(n\omega_{1\ell} + t\omega_{2\ell})} r_m(\psi_\ell) + w_{ntm},
\end{equation}
where $r_m(\psi_\ell)$ and $x_{n}(\varsigma_\ell)$ are defined in \eqref{eq:r_psi} and \eqref{eq:x_sigma}, respectively. In this case, the tensor $\mathcal{Y}$ again admits a \ac{CP}-structured decomposition. We first estimate the model order using the \ac{MDL} criterion on $\mathcal{Y}$, obtaining $\hat{L}$, and then compute the corresponding \ac{CP} decomposition, which yields components
\begin{align}
    &[\mathbf{a}^1_\ell]_n \: \propto \: b_\ell e^{jn\omega_{1\ell}} \sum_v e^{jv\varsigma_\ell} x_{nv}, \\
    &[\mathbf{a}^2_\ell]_t \: \propto \: e^{jt\omega_{2\ell}},\\
    &[\mathbf{a}^3_\ell]_m \: \propto \: \sum_u r_{mu} e^{ju\psi_\ell}.
\end{align}
To ensure adequate angular coverage, a $t$-constant doubly isotropic pilot-combiner pair can be designed as in \cite{hw_impaired_ch_est}, provided that $D_\text{T} \geq N^\text{T}_a$ and $D_\text{R} \geq N^\text{R}_a$.

The estimation then proceeds by first extracting $\omega_{2\ell}$ via \ac{ESPRIT}. Next, $\psi_\ell$ is obtained by solving
{\color{blue}
\begin{equation} \label{eq:psi_multistream_est}
    \hat{\psi}_\ell = \argmax_\psi \frac{\big|\sum_m [\mathbf{a}^3_\ell]^*_m \sum_u r_{mu} e^{ju\psi}\big|^2}{\sum_m \big|\sum_u r_{mu} e^{ju\psi}\big|^2},
\end{equation}}
which can be solved without numerical line search by using companion-matrix techniques such as those in \cite{ch_est_pinto}. The estimate of $\mathbf{a}^1_\ell$ is then rescaled\textcolor{blue}{, yielding $\hat{\mathbf{a}}^1_\ell$,} using the estimates of $\mathbf{a}^2_\ell$ and $\mathbf{a}^3_\ell$, following the same rationale as in Section \ref{sec:digital_estimation}. Finally, $\omega_{1\ell}$, $\varsigma_\ell$, and $b_\ell$ are obtained by first solving
\begin{equation} \label{eq:omega1_varsigma_est}
    (\hat{\omega}_{1\ell}, \hat{\varsigma}_\ell) = \argmax_{(\omega_{1}, \varsigma)} \frac{\left| \hat{\mathbf{a}}^{1,T}_\ell \boldsymbol{\beta}^*(\omega_1,\varsigma) \right|^2}{\sum_n \left|\sum_v e^{jv\varsigma} x_{nv} \right|^2 },
\end{equation}
where $[\boldsymbol{\beta}(\omega_1,\varsigma)]_n = e^{jn\omega_{1}} \sum_v e^{jv\varsigma} x_{nv}$, and then computing $b_\ell(\omega_{1}, \varsigma) = \hat{\mathbf{a}}^{1,T}_\ell \boldsymbol{\beta}^*(\omega_1,\varsigma)/ \| \boldsymbol{\beta}(\omega_1,\varsigma) \|^2$.


\vspace{-0.1cm}
\section{Algorithm Complexity}
We now analyze the computational complexity of the proposed method for the fully digital receiver case; for the hybrid combiner case, the same reasoning applies after replacing $N_\text{R}$ with $D_\text{R}$. In both the single-stream and multi-stream settings, the algorithm begins with model-order estimation followed by the \ac{CP} decomposition. Model-order estimation requires computing three \acp{SVD}, with complexity $\mathcal{O}(\max\{N_\text{C},N_\text{S},N_\text{R}\}\cdot N_x)$, where \mbox{$N_x = N_\text{C} N_\text{S} N_\text{R}$}. This first stage is not merely a preprocessing step: it determines the effective tensor rank used in the subsequent multilinear factorization and therefore directly influences the overall cost of the procedure.
The \ac{CP} decomposition is computed using the inexact \ac{GN-DL} algorithm, which has per-iteration complexity dominated by $\mathcal{O}(\hat{L} N_x)$ gradient computations
(given in \textit{flops}/iteration) plus $\mathcal{O}(\hat{L}^2 (N_\text{C} + N_\text{S} +N_\text{R}))$ Gramian-vector products \cite[Table 4.2]{cp_computation_algs}.

The remaining computations can be separated into $\hat{L}$ independent branches, each associated with one estimated \ac{CP} component. This structure is attractive from an implementation perspective since it enables parallel processing after the multilinear decomposition stage. Each branch performs one \ac{SVD} for each \ac{ESPRIT} step, followed by an \ac{ACD} procedure. The two \ac{ESPRIT} steps have combined complexity of $\mathcal{O}((N_\text{C} + N_\text{R}) \hat{L}^2)$. The \ac{ACD} step can be implemented using the exact line search strategy from \cite{ch_est_pinto}, which requires two \acp{EVD} per iteration, with complexity $\mathcal{O}(\max\{N_\text{T}, N_\text{S}\}^3)$ for the digital case and $\mathcal{O}(\max\{N_\text{T}, N_\text{C}\}^3)$ for the hybrid combiner case.

Overall, the computational cost is dominated by \ac{ACD} steps, so the total complexity of the proposed algorithm scales as $\mathcal{O}(\max\{N_\text{T}, N_\text{S}\}^3)$ ($\mathcal{O}(\max\{N_\text{T}, N_\text{C}\}^3)$ for the hybrid combiner case). Our analysis corroborates the design objective of the proposed tensor PCE method: 
to leverage structured tensor factorization to efficiently separate the components, then exploit lower-dimensional refinements and parallelizable updates to keep the remaining stages manageable.

\secgap
\section{Numerical Results}
We compare the proposed \ac{CREL-CP} algorithm with two representative \ac{SOTA} baselines: \ac{MS-SAGE} from \cite{hw_impaired_ch_est} and \ac{CP-TVFS} from \cite{cp-tvfs}. The \ac{MS-SAGE} algorithm can be viewed as a robust extension of the standard \ac{SAGE} procedure, with initialization strengthened through a multiple-start heuristic. \textcolor{blue}{The maximum number of paths for MS-SAGE is set to $L$ exactly}. For the tensor decompositions, we use Tensorlab 3.0 \cite{tensorlab}, with \ac{GN-DL} and \ac{ALS} for \ac{CREL-CP} and \ac{CP-TVFS}, respectively.

The \ac{MC} setup considers $N_\text{T} = N_\text{R} = 16$ antennas with fully digital receiver \textcolor{blue}{and hybrid beamformer sub-panels of size $N_a=2$}. We transmit an isotropic \cite{ch_est_pinto} single-stream pilot \textcolor{blue}{satisfying the conditions in Section~\ref{subsec:model_ss} for the digital case and use a multi-stream doubly isotropic pilot-combiner pair \cite{hw_impaired_ch_est} for the hybrid beamformer case}. In addition, \ac{CP-TVFS} performs numerical line search with a resolution of 4000 points. The pilot frame contains $N_\text{S}=64$ symbols and $N_\text{C}=31$ subcarriers \textcolor{blue}{with $f_{\rm scs}=30$~kHz spacing}. \textcolor{blue}{The noise variance is $N_0=\eta F f_{\mathrm{scs}}$, with thermal noise density $\eta=k_B T=-173$~dBm/Hz ($T=290$~K), and receiver noise figure $F=10$~dB.} The channel consists of $L=10$ paths, whose parameters $(\psi_\ell, \varsigma_\ell, \omega_{1\ell})$ are sampled from the three-dimensional uniform distribution $\mathcal{U}^3(-\pi,\pi)$. The Doppler angular frequency $\omega_{2\ell}$ is sampled from $\mathcal{U}(-\pi/16,\pi/16)$, corresponding to a maximum velocity range of approximately 67--145~km/h for a 7--15~GHz carrier. The path coefficient magnitudes $|b_\ell|$ follow a Rician distribution with noncentrality parameter $10^{-6}$ and scale parameter $5\cdot10^{-6}$, and \textcolor{blue}{the magnitude of the strongest path is multiplied by 10} to emulate a dominant \ac{LOS} component. Finally, the phases $\angle b_\ell$ are sampled from $\mathcal{U}(-\pi,\pi)$, and \textcolor{blue}{512} \ac{MC} realizations are generated for each simulated point. The generated channels include \ac{ICI} \cite{ici_channel}, i.e., the estimators have model mismatch.

\begin{figure}[!t]
    \centering
    \includegraphics[width=1\linewidth]{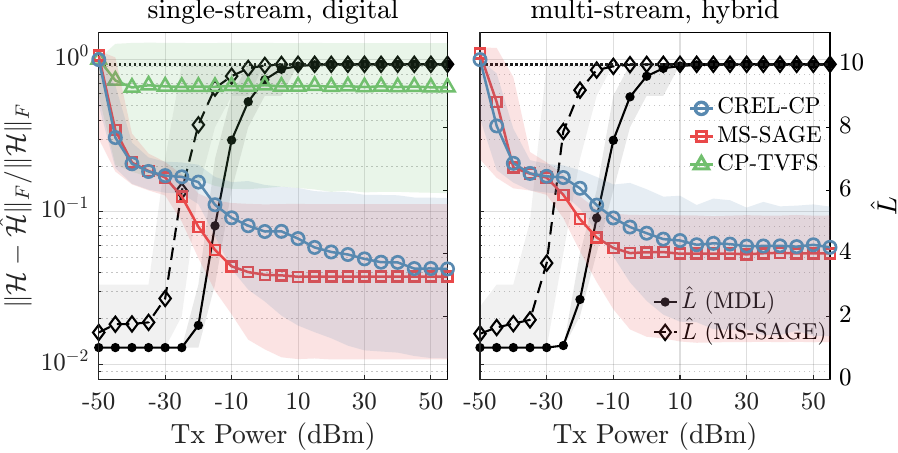} \vspace{-0.2cm}
    \caption{\textcolor{blue}{Channel tensor median relative estimation error (left axis) and model-order estimates (right axis) for the considered algorithms. The \nth{10} to \nth{90} percentiles are shaded. Left: single-stream, digital; right: multi-stream, hybrid.}}
    \label{fig:ch_est_model_order}
\end{figure}


In Fig.~\ref{fig:ch_est_model_order}, we show the channel tensor relative estimation error \mbox{$\|\mathcal{H} - \hat{\mathcal{H}}\|_F/\|\mathcal{H}\|_F$} \textcolor{blue}{(left axis)} and the model-order estimate $\hat{L}$ \textcolor{blue}{(right axis). \ac{CP-TVFS} has no multi-stream version.} As expected, the best estimation accuracy is achieved by \ac{MS-SAGE}, while \ac{CREL-CP} exhibits slightly lower accuracy. The alternating estimation of $\omega_{2\ell}$ and $b_\ell$ in \ac{CP-TVFS} yields poor accuracy and leads to large channel estimation errors. In addition, both \ac{CREL-CP} and \ac{CP-TVFS} rely on the \ac{MDL} criterion for estimating $\hat{L}$, since \cite{cp-tvfs} does not provide a dedicated model-order estimation procedure. Both \ac{MS-SAGE} and \ac{CREL-CP} have performance floors from the \ac{ICI} model mismatch.

\begin{figure}[!t]
    \centering
    \includegraphics[width=0.9\linewidth]{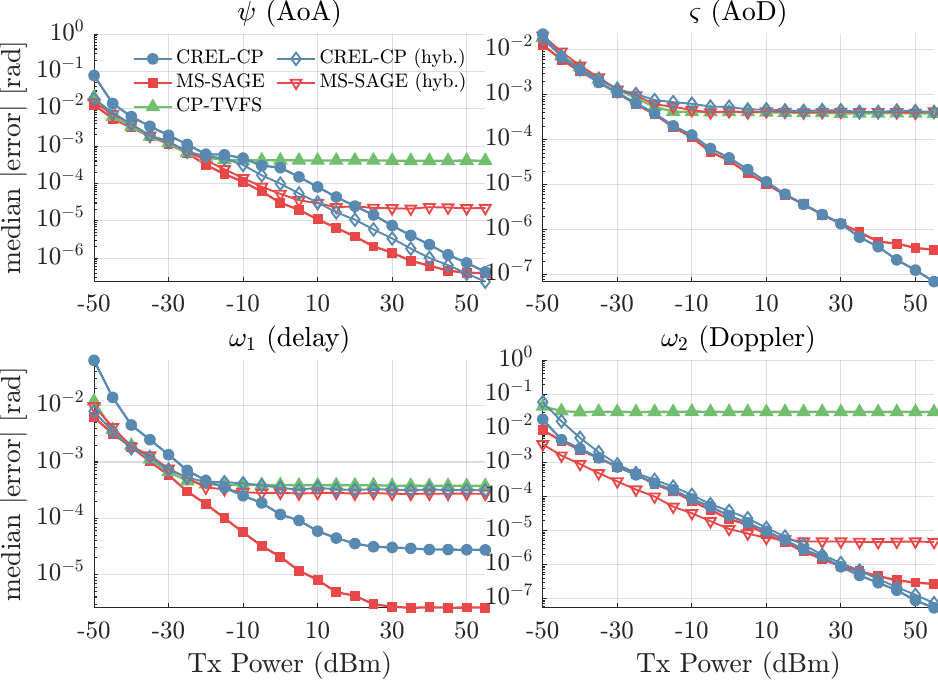} \vspace{-0.45cm}
    \caption{\ac{LOS} path median absolute error for each angular parameter\textcolor{blue}{, digital (filled markers) and hybrid (hollow markers)}.}
    \label{fig:los_param_error}
    \vspace{-0.15cm}
\end{figure}

Fig.~\ref {fig:los_param_error} shows the individual \ac{LOS} path parameter estimation performance of each method. As mentioned, \ac{CP-TVFS} has poor $\omega_{2\ell}$ estimation, which substantially degrades channel estimation. Furthermore, it has a performance floor due to the grid resolution\textcolor{blue}{, whereas the floors of \ac{CREL-CP} and \ac{MS-SAGE} are caused by \ac{ICI}. In the hybrid case, \ac{ICI} mixes symbols from neighboring subcarriers, compromising pilot-stream orthogonality and flooring the $\omega_{1\ell}$ and $\varsigma_\ell$ estimates at high \ac{Tx} power.}

Table~\ref {tab:exec_time} reports the wall-clock execution times at 30~dBm (representative of the general timing behavior of the algorithms above $10$~dBm) for both serial and parallel implementations. The results show that parallel \ac{CREL-CP} and \ac{CP-TVFS} are roughly \textcolor{blue}{$3$} times faster than \ac{MS-SAGE} \textcolor{blue}{in the single-stream case ($8$ times for \ac{CREL-CP} in the multi-stream case)}, which confirms the computational advantage of tensor-based processing over more demanding \ac{ML}-type iterative approaches. More importantly, \ac{CREL-CP} achieves this runtime reduction while maintaining a substantially better estimation quality than \ac{CP-TVFS}. These results corroborate the favorable tradeoff between estimation accuracy and computational efficiency offered by the \ac{CREL-CP} algorithm.

\begin{table}[!t]
    \centering
    \caption{Mean execution time over $64$ realizations at $30$\,\textnormal{dBm}.}
    \label{tab:exec_time}
    \footnotesize\color{blue}
    \setlength{\tabcolsep}{2.5pt}
    \begin{tabular}{lccccc}
    \hline
     & \multicolumn{3}{c}{Single-stream, digital} & \multicolumn{2}{c}{Multi-stream, hybrid} \\
     & CREL-CP & MS-SAGE & CP-TVFS & CREL-CP & MS-SAGE \\
    \hline
    Serial [s]       & $24.0$ & $28.5$ & $31.1$ & $8.0$ & $48.7$ \\
    Parallelized [s] & $\phantom{0}5.2$ & $16.5$ & $\phantom{0}4.6$ & $4.4$ & $34.9$ \\
    \hline
    \end{tabular}
\end{table}

\secgap
\section{Conclusion}
In this letter, we introduced the \ac{CREL-CP} algorithm, a tensor-based \ac{PCE} method that combines \ac{CP} decomposition, component rescaling, and efficient parameter updates to address fast time-varying channels. The proposed approach improves upon a related tensor-based \ac{SOTA} baseline while remaining significantly less computationally demanding than a \ac{SAGE}-based \ac{SOTA} solution. Although the latter still provides slightly more accurate estimates, \ac{CREL-CP} achieves a much more attractive complexity-performance tradeoff and has much lower execution time. Our results indicate that \ac{CREL-CP} is a fast, effective alternative to more expensive \ac{EM}-based \ac{PCE} algorithms in high-mobility scenarios.

\vspace{-0.2cm}

\bibliographystyle{ieeetr}
\bibliography{biblio}

\end{document}